\documentclass[12pt]{iopart}

\usepackage{graphicx,color,ulem}

\usepackage{hyperref}
\hypersetup{
    colorlinks=true,       
    linkcolor=blue,          
    citecolor=blue,        
    filecolor=blue,      
    urlcolor=blue           
}

\usepackage{doi}

\graphicspath{{./}{./figs/}}

\begin{document}

\title[Minimally destructive, Doppler measurement of a quantized flow in a ring BEC]{Minimally destructive, Doppler measurement of a quantized flow in a ring-shaped Bose-Einstein condensate}

\author{A. Kumar, N. Anderson, W.D. Phillips, S. Eckel, and G.K. Campbell}
\address{Joint Quantum Institute, National Institute of Standards and Technology and University of Maryland, Gaithersburg, Maryland 20899, USA}
\author{S. Stringari}
\address{INO-CNR BEC Center and Dipartimento di Fisica, Universit\`{a} di Trento, 38123 Povo, Italy}

\begin{abstract}
The Doppler effect, the shift in the frequency of sound due to motion, is present in both classical gases and quantum superfluids.  Here, we perform an {\it in-situ}, minimally destructive measurement,  of the persistent current in a ring-shaped, superfluid Bose-Einstein condensate using the Doppler effect.  Phonon modes generated in this condensate have their frequencies Doppler shifted by a persistent current.  This frequency shift will cause a standing-wave phonon mode to be ``dragged'' along with the persistent current.   By measuring this precession, one can extract the background flow velocity.  This technique will find utility in experiments where the winding number is important, such as in emerging `atomtronic' devices.\end{abstract}

\maketitle

\section{Introduction}


Ring-shaped Bose-Einstein condensates (BECs) use topology to exploit one of the key features of a BEC: superfluidity.  In particular, the topology supports superfluid persistent currents~\cite{Ryu2007}.  As a result, a ring-shaped condensate forms the basis of several so-called `atomtronic' devices: simple circuits that resemble counterparts in electronics~\cite{Ramanathan2011,Wright2013,Eckel2014,Ryu2013,Jendrzejewski2014}.
The addition of one or more rotating perturbations or weak links into the ring can form devices that are similar to the rf-superconducting quantum interference device (SQUID)~\cite{Ramanathan2011,Wright2013,Eckel2014} and dc SQUID~\cite{Ryu2013,Jendrzejewski2014}.  Operation of these devices typically requires measuring the persistent current.  Here, we present a technique for measuring the persistent current of a ring that uses the Doppler effect and, unlike other methods, is done {\it in-situ} and can be minimally destructive.  


Superfluids can be described using a macroscopic wavefunction (or order parameter) $\psi = \sqrt{n}e^{i\phi}$, where $n$ is the density of atoms and $\phi$ is phase of the wavefunction.  In this picture, the flow velocity is given by $v = (\hbar/m)\nabla\phi$, where $m$ is the mass of the atoms and $\hbar$ is the reduced Planck's constant.  Because the wavefunction must be single valued, the integral $\oint \nabla\phi\cdot\ dl$ must equal $2\pi\ell$, where $\ell$ is an integer called the winding number.  This quantization of the winding number forces flows around a ring of radius $R$ to be quantized with an angular velocity $\Omega_0 = \hbar/mR^2$.  Thus, the angular flow velocity must satisfy $\Omega = \ell\Omega_0$.

\begin{figure}
	\center
	\includegraphics{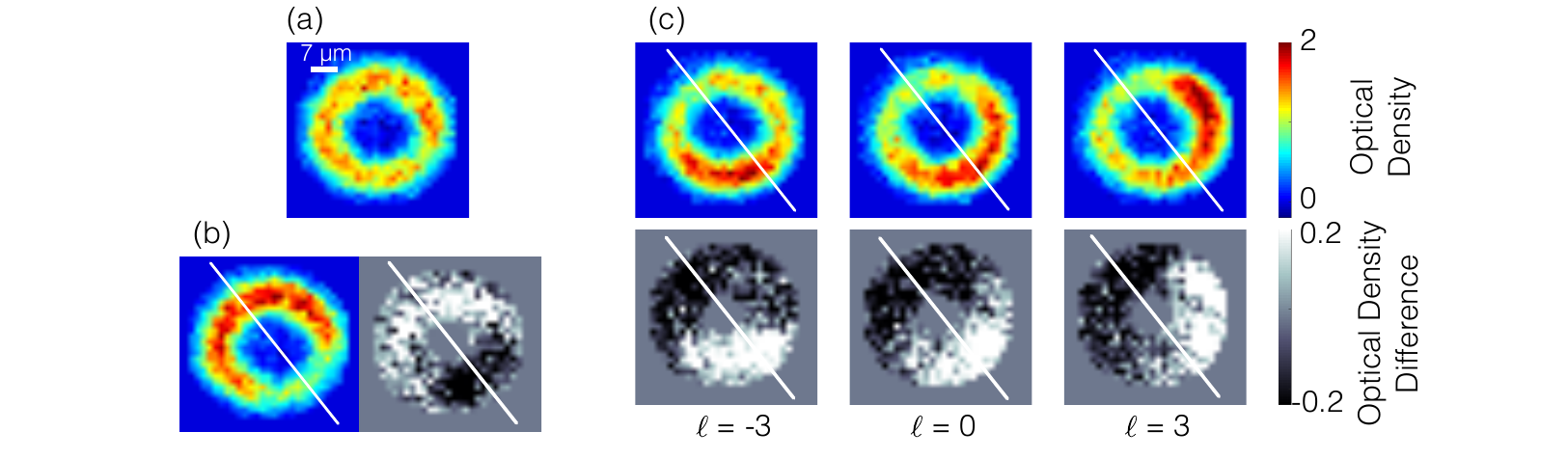}
	\caption{\label{fig:intro} (a) Image of the ring condensate {\it in-situ} without an applied perturbation.  (b) A sinusoidal perturbation excites a standing-wave superposition of counterpropogating phonon modes.  The white line intersects the ring at the maximum and minimum of the perturbation.  The left image shows the density; the average density is subtracted from the right image.  These images show the resulting density modulation, $100$~$\mu$s after the perturbation has been removed.  (c) Full density (top) and average-subtracted-density (bottom) images taken 9.1~ms after removal of the perturbation.  The density modulation rotates relative to the initial perturbation in the presence of superfluid flow.  The winding numbers are shown below the images.} 
\end{figure}

In addition to supporting bulk persistent flows, the BEC also serves as a medium in which sound can travel.  Because of the trap's boundary conditions, only certain wavelengths of sound are permitted.  These phonon, or Bogoluibov, modes are the lowest energy collective excitations of the condensate~\cite{Stringari1996,Edwards1996}.  In this work, we excite a standing wave mode with wavelength $\lambda=2\pi R$, as shown in Fig.~\ref{fig:intro}(a-b).  Such a standing wave is an equal superposition of clockwise and counterclockwise traveling waves with the same wavelength.  In the presence of a background flow, the Doppler effect shifts the relative frequencies of the two traveling waves.  This shift causes the standing wave to precess, as shown in Fig.~\ref{fig:intro}(c).  Here, we use this precession to detect the background flow velocity of the superfluid.  This method is analogous to earlier techniques, where the precession of a quadrupole oscillation was used to measure~\cite{Chevy2000,Haljan2001} the sign and charge of a quantized vortex in a simply connected, harmonic trap, as suggested in Refs.~\cite{Zambelli1998,Svidzinsky1998}.  

In a ring-shaped condensate, two other methods have been used to measure the background flow velocity by determine the winding number $\ell$.  Both are inherently destructive because they require that the BEC be released from the trap.  First, sufficient expansion of the condensate can yield a hole at the center of the cloud whose size is quantized according to $\ell$~\cite{Cozzini2006,Moulder2012,Murray2013,Ryu2014}.  Second, experiments releasing both a ring and a reference condensate produce spiral interference patterns that indicate $\ell$~\cite{Eckel2014b,Corman2014}.

Our method of detecting rotation by the Doppler effect is sufficiently precise to distinguish adjacent values of $\ell$.  However, it is minimally disruptive: it requires only exciting a sound wave and imaging the resulting density modulation.  By imaging the density modulation using a minimally destructive imaging method (such as dark-field dispersion imaging~\cite{Andrews1996}, phase contrast imaging~\cite{Andrews1997}, diffraction-contrast imaging~\cite{Turner2005}, partial transfer absorption imaging~\cite{Ramanathan2012}, or Faraday imaging~\cite{Gajdacz2013}) one can create a winding number measurement that is also minimally destructive.  Such measurements could allow easier implementation of experiments where rings are stirred by weak links more than once, such as hysteresis experiments.  By measuring the winding number in a minimally destructive way, the quantum state of the ring can be known after each stage of stirring.   One can use multiple minimally-destructive winding number measurements to increase the sensitivity of a rotation sensor based on the atomtronic rf-SQUID.  Our system~\cite{Eckel2014} has a sensitivity of $\approx0.1$~Hz for a measurement done once every $\approx30$~s, which includes creating the condensate, stirring, and measuring the winding number.  This leads to an effective sensitivity to rotation of $\sim5$~Hz/$\sqrt{\rm Hz}$.  By repeating stirring and winding number measurement on a single condensate using a minimally-destructive technique (which yields the same information as a destructive measurement), the sensitivity improves by $\sqrt{N_{\rm meas}}$, where $N_{\rm meas}\approx10$ is an achievable number of minimally-destructive measurements.


Finally, it is possible to use this technique to detect rotations of the inertial frame of the condensate itself, using the Sagnac effect\footnote{\label{foot:sagnac} We note that unlike the ring-shaped condensate considered here, not every mode in a simply connected condensate is sensitive to both a rotating frame and to vorticity.  In particular, the dipole oscillation mode is sensitive to a rotating frame but not sensitive to vorticity~\cite{Zambelli1998}.}.  Ref.~\cite{Marti2015} measured the noise level associated with such an excitation-based rotation sensor, finding a sensitivity in their system of roughly 1~(rad/s)/$\sqrt{\rm Hz}$.  However, their experiment was not configured to produce rotation, so the effect was not measured experimentally.  Here, we demonstrate the feasibility of this idea, by detecting the quantized rotation of the superfluid in the ring itself.

\section{Theory}

To understand the Doppler effect and its effect on phonons in the ring, let us first consider a standing-wave phonon mode in the absence of a persistent current.  To simplify the calculation, let us consider the equivalent problem of a 1D gas of length $2\pi R$, where $R$ is the radius of the ring, and let us impose periodic boundary conditions.  A standing wave mode can be generated by a perturbation of the form $V = V_0\cos(q\theta)$, where $q$ is a positive integer, $V_0$ is the amplitude of the perturbation, and $\theta$ is the azimuthal angle.  This perturbation generates a modulation in the density of the form $n_{\rm 1D}(\theta) = n_{\rm eq} + \delta n\cos(q\theta)$, where $n_{\rm eq}$ is the equilibrium density without the perturbation and $\delta n$ is the amplitude of the density modulation.  Upon sudden removal of the perturbation, the density $n_{\rm 1D}(\theta)$ is projected onto the spectrum of phonon modes.  The initial density modulation is then described by the superposition of two counterpropogating modes $\cos(\omega(q_\pm) t - q_\pm \theta)$.  Here, $q_+$ indicates a mode traveling counterclockwise and $q_-$ indicates a mode traveling clockwise.    In the absence of a persistent current the two modes have equal frequency: $\omega(q_\pm)=\omega_0$.  Consequently, the density modulation will oscillate in time, $\delta n(t) \propto \cos(\omega_0 t)\cos(q\theta)$, without exhibiting precession.

The presence of the current will remove the degeneracy between the two modes through the Doppler effect.  The frequency of the two modes, in the presence of a superflow of velocity $v=\ell\Omega_0R$, will be given by
\begin{equation}
	\label{eq:shift}
	\omega(q_\pm) = \omega_0 \pm \frac{q}{R} v\ .
\end{equation}
The density modulation is then proportional to
\begin{eqnarray}
	& & \cos[q_+\theta-\omega(q_+)t] + \cos[q_-\theta+\omega(q_-)t] \nonumber \\
	           & & ~~~~~~~ = \cos\left[q\left(\theta - \frac{v}{R} t\right)\right]\cos(\omega_0 t)
\end{eqnarray}
and hence evolves in space.  The azimuthal location $\phi_m$ of an anti-node of this density standing wave precesses around the ring as
\begin{equation}
	\label{eq:main_result}
	\frac{d\phi_m}{dt} = \frac{\ell \hbar}{m R^2}\ .
\end{equation}
Measurement of $\phi_m(t)$ would consequently provide a direct measurement of $\ell$.  We note that $d\phi_m/dt$ does not depend on $\omega_0$; therefore, the speed of sound and the details of the phonon dispersion curve are not relevant in predicting the precession.

While the above arguments may seem to apply only to a ring, they can easily be generalized to any quasi-one-dimensional geometry.  In such a case, the shifts in the phonon frequencies will still be given by Eq.~\ref{eq:shift}, but with a $v$ that may or may not be quantized.  For an infinite one-dimensional channel, for example, the flow is not quantized.  However, two equal but oppositely directed phonon wavepackets (similar to those generated in Ref.~\cite{Wang2015}) traveling in the channel will move with different velocities in the presence of a background flow.  The use of  a ring allows for a straightforward means of detecting the frequency shift: precession of a phonon standing-wave mode.

\section{Experimental details}

\begin{figure}
	\center
	\includegraphics{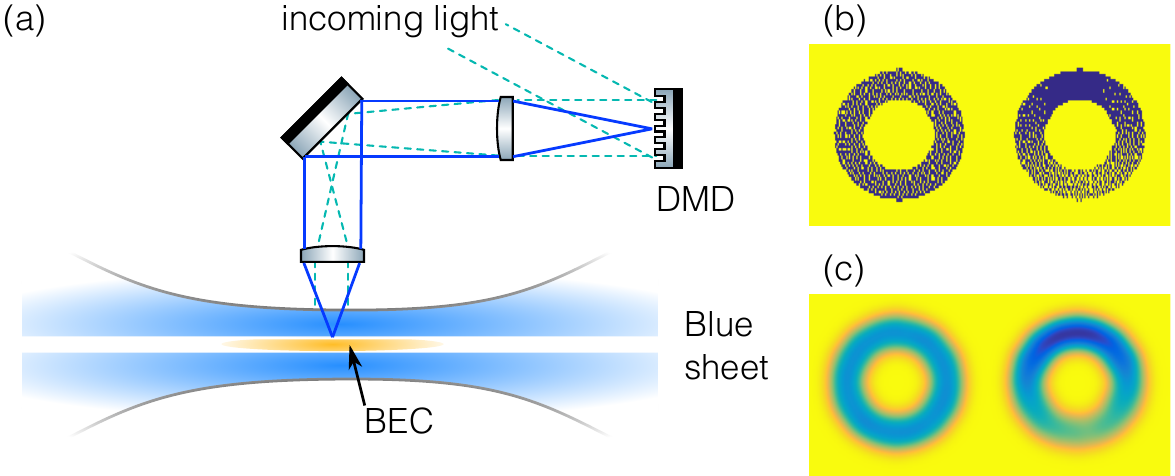}
	\caption{\label{fig:setup} (a) Schematic of the setup used to trap the atoms and create the double ring potential and sinudoidal perturbation.  A DMD is illuminated with blue-detuned light and is imaged onto the atoms using a telescope (solid lines) , which ensures that incoming collimated beams remain collimated after they emerge from the telescope (dashed lines).  (b) The halftoned patterns~\hyperref[foot:halftone]{$\|$} written to the DMD for the bare ring potential (left) and the perturbed ring potential (right).  (c) After convolution with the point-spread function of the imaging system, the potentials formed are smooth, and show a clear sinusoidal perturbation (right).}
\end{figure} 

We create a $^{23}$Na condensate in a crossed-optical dipole trap, as shown in Fig.~\ref{fig:setup}a.  A blue-detuned double sheet beam, formed by focusing a TEM$_{01}$ mode tightly along the vertical direction, provides vertical confinement.  
Confinement in the horizontal plane is generated using another blue detuned beam that is shaped using a Texas Instruments LightCrafter 3000 digital micromirror device (DMD) \footnote{The identification of commercial products is for information only and does not imply recommendation or endorsement by the National Institute of Standards and Technology.}.  We position the DMD in our imaging system to directly image the surface of the DMD onto the atoms. (This method is similar to the photomask method used in Ref.~\cite{Lee2014}, with the DMD replacing the photomask.)   

The DMD generates a double ring trap.  These experiments use only the inner ring, which is shown in Fig.~\ref{fig:intro}(a).  In general, our condensates contain $\approx 7\times10^5$ atoms, with approximately $5\times10^5$ in the outer ring and $2\times10^5$ in the inner ring.  The outer ring, with mean radius 31(1)~$\mu$m~\footnote{Unless stated otherwise, uncertainties represent the 1$\sigma$ combination of statistical and systematic errors.}, can be used as a phase reference to measure the winding number of the inner ring~\cite{Mathew2015}.  The inner ring has a chemical potential of $\mu/h = 3.1(6)\mbox{ kHz}$.  The vertical trapping frequency is 1020(30)~Hz and the radial trapping frequency in the inner ring is 310(10)~Hz.

In order to generate a persistent current in our ring, we apply a rotating potential generated by a blue-detuned beam steered by an acoustic-optic deflector (AOD).  A previous paper~\cite{Eckel2014} describes this stirring technique.  We verify that this produces the desired $\ell$ state 95\% of the time by performing a fully-destructive interference measurement~\cite{Eckel2014b}.  The phase winding in the inner ring can be measured by interfering the inner ring with the outer ring condensate, which serves as a phase reference.  Fig.~\ref{fig:setup}(d) shows a typical interference pattern, which indicates $\ell=-1$.  (We note that the chirality of these spirals for a given sign of the winding number is opposite of those in Ref.~\cite{Eckel2014b}, because the reference condensate is on the outside rather than the inside of the ring.)

In addition to generating the static trap, the DMD can also produce perturbations to the potential.  However, because any individual mirror is binary (either on or off), potentials that require intermediate values require a form of grayscale control.  We achieve such control by using Jarvis halftoning~\cite{Jarvis1976}.  Because the point-spread function of our imaging system (in the plane of the atoms, $\approx6$~$\mu$m $1/e^2$ full-width) is much larger than the DMD pixel size (in the plane of the atoms, $\approx 0.5$~$\mu$m), the potential at any given location is the convolution of the binary values of all the nearby pixels with the point spread function.  For example, Fig.~\ref{fig:setup}(b) shows a halftoned pattern written to the DMD for generating a sinusoidal perturbation of the form $V_0\sin(\theta)$~\footnote{\label{foot:halftone}The halftoning is more evident if one zooms in on Fig.~2b.}.  Convolution with the point-spread function generates the desired sinusoidal potential, as shown in Fig.~\ref{fig:setup}(c).

We empirically find that a perturbation of $V_0\approx0.4\mu$ applied for 2~ms is sufficient to excite the first phonon mode without perturbing the flow state of the ring.   The speed with which our perturbation can be turned on and off is limited by the refresh rate of our DMD to 250~$\mu$s.  To image the resulting density modulation {\it in-situ}, we use partial transfer absorption imaging (PTAI)~\cite{Ramanathan2012}, which is a type of minimally-destructive imaging~\cite{Hope2004}.  In general, we transfer approximately 5\% of the atoms into the imaging state.  Despite the method being minimally destructive, none of the experiments contained in this work use repeated imaging on the same condensate.

\section{Results}

\begin{figure}
	\center
	\includegraphics{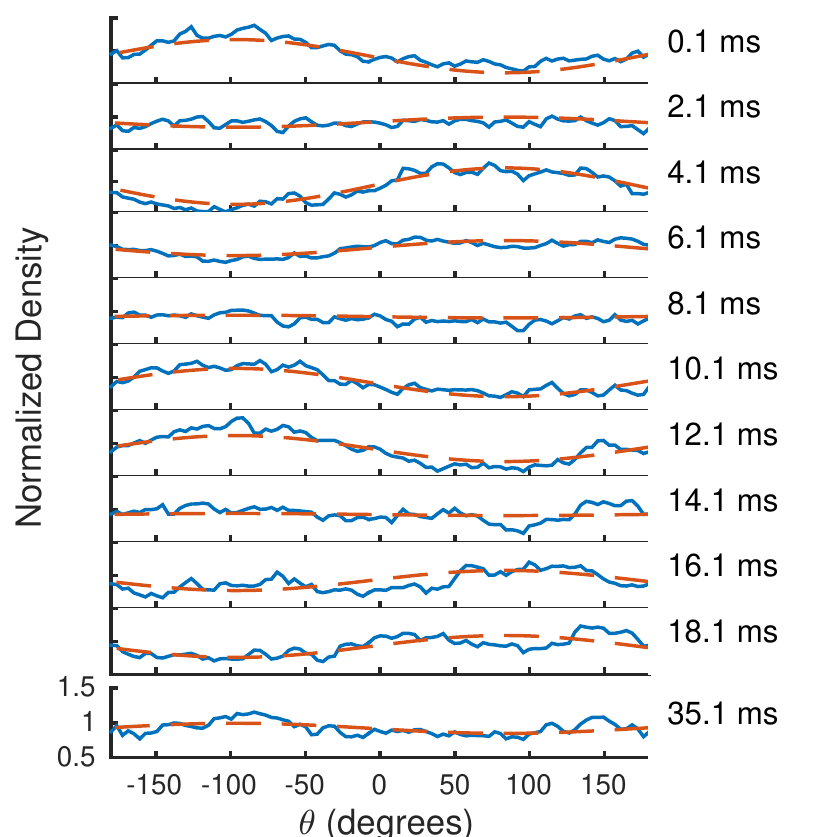}
	\caption{\label{fig:time_evolution_cutthroughs} Plot of the normalized 1D density of the ring $n_{\rm 1D}(t,\theta)/n_{\rm 0,1D}(\theta)$ (see text) vs. angle, for different times after the perturbation.  The solid blue lines show the experimental data and the red dashed lines show the fit to the data.  Each trace represents one shot of the experiment.  For these data, there is no persistent current present ($\ell=0$).}
\end{figure}

To detect rotation accurately in the ring, the behavior of the phonon mode in the absence of a persistent current must first be understood.  To this end, we apply our perturbation to the inner ring and observe the resulting oscillation in the phonon mode, without a persistent current ($\ell=0$).   Shown in Fig.~\ref{fig:time_evolution_cutthroughs} is the normalized 1D density $n_{\rm 1D}(t,\theta)/n_{\rm 0,1D}(\theta)$, where $n_{\rm 1D}(t,\theta)$ is the 1D density measured a time $t$ after the perturbation was applied and $n_{\rm 0,1D}(\theta)$ is the 1D density separately measured with no perturbation applied.  The density $n_{\rm 1D}$ is determined via $\int n_{\rm 2D}(t,r,\theta)\ rdr$, where the 2D density $n_{\rm 2D}(t,r,\theta)$ is determined from imaging followed by interpolated conversion into polar coordinates.  The Bogoliubov mode oscillates with time in the ring, but is also damped (as discussed below).  At $\theta\approx\pm90^\circ$, maxima and minima appear immediately after the perturbation.  At these angles, clear oscillations are seen as a function of time.  At the nodal points $\theta\approx0^\circ,180^\circ$, oscillations are expected to be absent in a perfectly uniform ring.

\begin{figure}
	\center
	\includegraphics{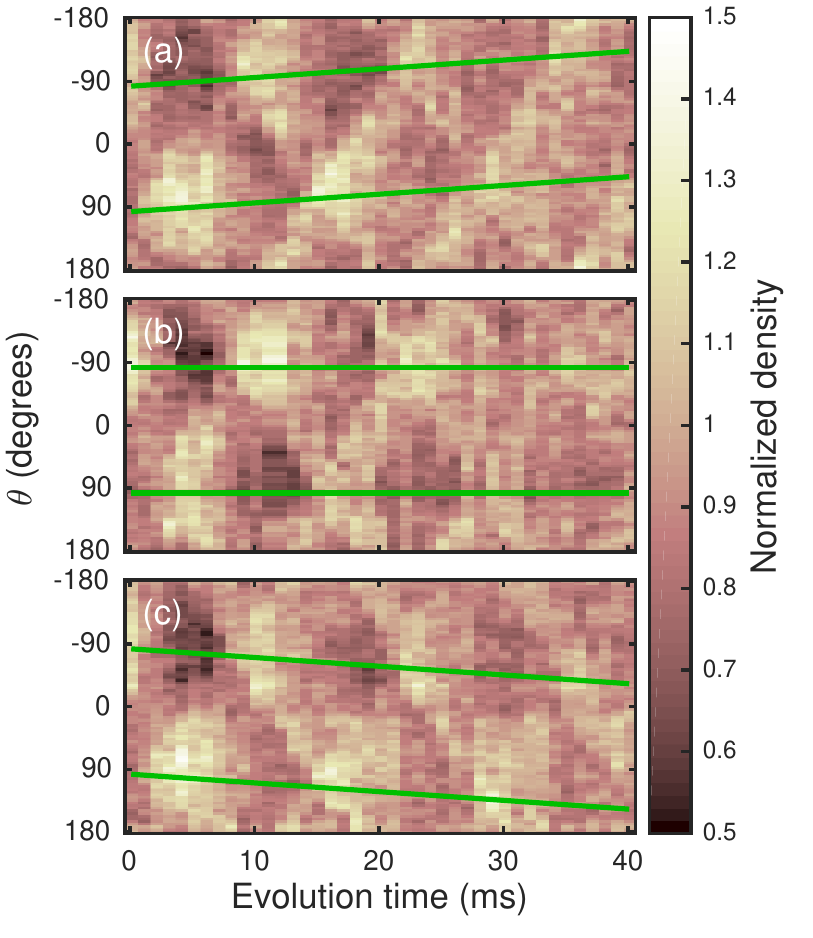}
	\caption{\label{fig:time_evolution} Plot of the normalized 1D density as a function of azimuthal angle and hold time after the perturbation.  The three panels correspond to having different persistent currents in the ring: (a) $\ell=-1$, (b) $\ell=0$, and (c) $\ell=1$.  For each time and winding number, there is one experimental shot (i.e., there is no averaging).  The green lines show the expected precession of the antinodes of the standing wave, according to Eq.~\ref{eq:main_result}.}
\end{figure}

We fit the data to the function $a e^{-t/\tau}\sin(\omega t+\phi_1)\sin(\theta+\phi_2)+c$, where $a$, $\tau$, $\omega$, $\phi_{1,2}$, and $c$ are fit parameters.  The average best fit value of the oscillation frequency is 79.3(3)~Hz.  We can estimate the frequency of this fundamental phonon mode as $\omega_0 =c_s/R$, where $c_s$ is the speed of sound.  Using the speed of sound for a narrow channel $c_s =\sqrt{\mu/2 m} = 5.3(5)$~mm/s~\cite{Zaremba1998} yields $\omega_0/2\pi=74(7)$~Hz, in good agreement with the average best fit value.  The average best fit value of the decay constant is $\tau = 25.6(1)$~ms, so that by 100~ms, the oscillation amplitude is diminished well below the noise.

The decay of oscillation could be caused by a number of effects, including Landau damping and Beliaev damping, i.e., four-wave mixing.  We expect Landau damping to thermalize our excitations by scattering from other, thermal, phonons~\cite{Fedichev1998}.  In this case, we expect that the quality factor $Q=\omega \tau$ will be independent of the mode number $q$~\cite{Marti2015}.  By contrast, damping via four-wave mixing, called Beliaev damping, will result in a strong dependence of $Q$ on mode number~\cite{Katz2002,Marti2015}.  Landau damping should also depend on the temperature, while Belieav damping should depend on the density, but we have not explored these dependencies~\footnote{Furthermore, it is not clear if the phonon spectrum in one-dimension will fulfill momentum conservation in a four-wave mixing process.}.  Nevertheless, we observe a $Q$ for the $q=\pm 1$ mode superposition of $12.8(6)$.  We also investigated the decay of the $q=\pm 2$ mode, which has a $Q$ of $13.8(4)$, agreeing within the uncertainties with the $Q$ for the decay of the $\pm 1$.  Therefore, Landau damping appears to be favored over Beliaev damping as the dominant damping mechanism.

Having studied the relevant features of the phonon mode without a persistent current, we can now use Eq.~\ref{eq:main_result} to predict the precession of the anti-nodes and nodes in the presence of a persistent current.  The antinode initially at $\theta=-90^\circ$ reaches maximum density at approximately 10, 24, and 35~ms; the antinode initially at $\theta=90^\circ$ reaches maximum density at approximately 5, 17, and 29~ms (see Fig.~\ref{fig:time_evolution})~\footnote{The maximum contrast of the standing wave does not occur at time $t=0$ because of the details of how the perturbation is applied.}.  At these times, the location of these antinodes is determined by Eq.~\ref{eq:main_result}.  Shown in Fig.~\ref{fig:time_evolution}(a), (b), and (c) are the cases where the winding number in the ring prior to the perturbation was $\ell=+1$, $0$, and $-1$, respectively.  Given the radius of the ring, we expect that these peaks should move at rate of $\ell\hbar/mR^2 t = [\ell \times 1.25(7)\mbox{ deg/ms}] t$.  This rate is shown by the green lines in Fig.~\ref{fig:time_evolution}, and follows the peaks well.

\begin{figure}
	\center
	\includegraphics{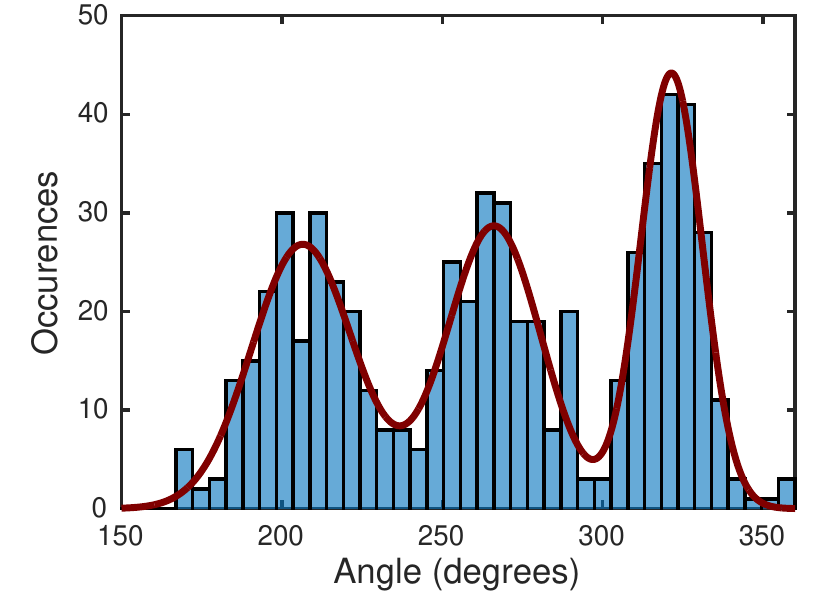}
	\caption{\label{fig:histogram} Histogram of the number of occurrences of measured precession angles of one of the antinodes of the standing wave at $t=35.1$~ms.  The red curve is a fit of three, independent Gaussians.  These Gaussians correspond to three different winding numbers: $\ell=-1$, $\ell=0$, and $\ell=1$ (left to right).}
\end{figure}

To detect a flow velocity with sufficient precision to distinguish between adjacent winding numbers using a single image, one needs to balance the observation time with the signal to noise ratio.  While the angular deflection grows linearly with time, the amplitude of the oscillation, and thus its signal to noise, decreases roughly exponentially with time.  Empirically, we find that time $t=35$~ms is the best compromise; here, the difference in deflection for $\Delta\ell=1$ is expected to be $\approx 45^\circ$. 
To test our ability to determine a given winding number with a single image, we fit the maximum of the density profile for 614 individual repetitions initialized with winding numbers $\ell=0$, $-1$ or $1$.  We bin the results to form a histogram, shown in Fig.~\ref{fig:histogram}.   The histogram shows three clear peaks, which we fit to Gaussians.  Each corresponds to one of the three winding numbers.  Each has a slightly different width because of the compression or expansion of the phonon wave in the azimuthally varying density profile of the ring.

We can use the Gaussian fits to predict the confidence with which we can assign a winding number.  Given the strong overlap of adjacent Gaussians, we have a confidence of $\approx90\%$ of identifying a $\ell=-1$, and a confidence of $\approx 95\%$ of identifying an $\ell=1$ state.  These confidences can be made better by attempting to make the ring more uniform to keep the form of the oscillation well defined.  This might improve the fitting algorithm and decrease the noise.  In addition, if the condensate were colder, the damping of the phonon mode should be reduced.  This would allow for longer interrogation times and therefore more angular displacement between adjacent $\ell$ states.

For this this method of measuring winding number to be minimally destructive, one must use a minimally destructive imaging method~\cite{Hope2004}.  Previous experiments~\cite{Freilich2010} show that such imaging methods do not perturb the flow state of BECs. In addition, the perturbation one applies to the condensate must also be minimally disruptive.  The observed fast decay helps to ensure this.  Perhaps most important, the application of the perturbation must not change the winding number.  To verify this, we measured the winding number that results from stirring at multiple different rotation rates using our destructive interference technique.  (This produced data similar to those in Ref.~\cite{Wright2013}.)   We then repeated this experiment, but after stirring, applied our sinusoidal perturbation, allowed 100~ms for the ring to settle, and measured the winding number again.  The data show no significant change in the winding number as a result of the application of the perturbation.

One might ask if going to higher modes, particularly $q=\pm2$, would yield better sensitivity to rotation.  The frequency of the oscillation increases as $q$, and because the quality factors of the modes are the same, the time at which one achieves a given signal-to-noise ratio scales as $t\sim q^{-1}$.  Because the precession angle scales as $\sim t$, these two factors cancel.  However, one might be able to better determine the angular position of the maxima because the width of the oscillation peaks decreases as $q^{-1}$.  Further work is needed to verify these scalings.

\section{Conclusion}

We have demonstrated a minimally destructive {\it in-situ} method of measuring winding number in a ring-shaped Bose-Einstein condensate.  This technique can be used as an excitation-based Sagnac interferometer~\cite{Marti2015}~\hyperref[foot:sagnac]{$\dagger$}.  Because of its non-destructive behavior, this method can be applied to a variety of ring experiments, and may possibly be used to increase the sensitivity of atomtronic rotation sensors, by being able to repeat the measurement of winding number many times on a single condensate.

\section*{Acknowledgments}
This work was partially supported by ONR, the ARO atomtronics MURI, and the NSF through the PFC at the JQI.  S.S. acknowledges the support of the QUIC grant of the Horizon 2020 FET program and of Provincia Autonoma di Trento.

\section*{References}
\bibliographystyle{iopart-num}
\bibliography{insitu_winding_library}

\end{document}